\documentclass[12pt]{article}

\def\bfkappa{\mathop{\mbox{\boldmath $\kappa$}}}

\newcommand{\be}{\begin{equation}}
\newcommand{\ee}{\end{equation}}
\newcommand{\bi}[1]{\vspace{-3mm} \bibitem{#1}}

\usepackage{epsfig}
\usepackage{epsfig,amsmath}
\usepackage{graphics}

\begin{document}

\begin{center}
{\it Physica A 354 (2005) 249-261}
\vskip 5 mm

{\Large \bf Fractional Ginzburg-Landau equation \\
for fractal media}
\vskip 5 mm

{\large \bf  Vasily E. Tarasov$^{1}$ and George M. Zaslavsky$^{2,3}$ }\\

\vskip 3mm

{\it $1)$ Skobeltsyn Institute of Nuclear Physics, \\
Moscow State University, Moscow 119992, Russia } \\
{\it $2)$ Courant Institute of Mathematical Sciences, New York University \\
251 Mercer St., New York, NY 10012, USA, }\\ 
{\it $3)$ Department of Physics, New York University, \\
2-4 Washington Place, New York, NY 10003, USA } \\
\end{center}

\vskip 11 mm

\begin{abstract}
We derive the fractional generalization of the Ginzburg-Landau equation
from the variational Euler-Lagrange equation for fractal media.
To describe fractal media we use the fractional integrals
considered as approximations of integrals on fractals.
Some simple solutions of the Ginzburg-Landau equation
for fractal media are considered and different forms of the 
fractional Ginzburg-Landau equation or nonlinear 
Schr\"{o}dinger equation with fractional derivatives 
are presented. The Agrawal variational 
principle and its generalization have been applied.
\end{abstract}

\vskip 7 mm

PACS:  03.40.-t; 05.45.Df; 47.53.+n  \\


\section{Introduction}

We call the fractional equation an equation 
that contains fractional derivatives or integrals. 
Derivatives and integrals of fractional order have found many
applications in recent studies in physics.
The interest in fractional equations has been growing continually 
during the last few years because of numerous applications: 
kinetic theories of systems with chaotic dynamics 
\cite{Zaslavsky2,Zaslavsky8};  
pseudochaotic dynamics \cite{Zaslavsky9}; 
dynamics in a complex or porous media \cite{Nig}; 
random walks with a memory and flights 
\cite{Shlesinger,Zaslavsky10,Zaslavsky4}; 
and many others.

Fractional integrals and derivatives can be used 
to describe processes with different levels of complexity. 
The new type of problems has rapidly increased interest in areas 
in which the fractal features of a process or the medium 
impose a necessity of applying tools that are non-traditional  
in "regular" smooth physical equations. 
In many problems the real fractal structure of matter
can be disregarded and the medium can be replaced by
some smoothed continuous mathematical model where
fractional integrals appear \cite{PLA05}.
The order of the fractional integral is equal
to the fractal mass dimension of the medium and in this way one can 
take into account the fractality of the media.
It was proved that integrals on a net of fractals can be approximated 
by fractional integrals \cite{RLWQ} and that fractional integrals
can be considered as integrals over the space with fractional
dimension up to a numerical factor \cite{chaos}. The last proof  used the 
formulas of dimensional regularizations \cite{Col}.

The fractional generalization of the Ginzburg-Landau equation
was suggested in Ref. \cite{Zaslavsky6}.
This equation can be used to describe the dynamical processes in a media with
fractal dispersion. Since the fractals can be realized in nature as 
a fractal process or fractal media, it is interesting to derive 
fractional Ginzburg-Landau (FGL) equation using a
corresponding generalization of the free energy functional.

It is known \cite{LP}  that the Ginzburg-Landau equation 
\[ g \Delta Z-aZ-bZ^3=0 \]
can be derived as the variational Euler-Lagrange equation
\be \label{1}  \frac{\delta F\{Z({\bf x})\}}{\delta Z ({\bf x})}=0  \ee
for the free energy functional
\be \label{2} F\{Z({\bf x})\}=F_0+\frac{1}{2}\int_W[g(\nabla Z)^2 +
aZ^2+\frac{b}{2}Z^4 ] dV_3 , \ee
where the integration is over three-dimensional region $W$ of
continuous media. Here $F_0$ is a free energy of the normal
state, i.e. $F\{Z\}$ for $Z=0$.
In this paper we consider a fractional generalization of (\ref{2})
that can appear from two places: fractional generalization 
of the integral in Eq. (\ref{2}) and  fractional generalization 
of the derivatives in Eq. (\ref{2}). Different forms of
the FGL equation are considered in the coordinate space and 
in the dual space after applying the Fourier transform.   
Finally, the Agrawal variational principle \cite{Agrawal} and 
its generalization have been applied to obtain the FGL equation.

\section{Fractional generalization of free energy functional}

Let us consider the thermodynamic potential (free energy
functional) $F\{Z({\bf x})\}$ for the non-equilibrium state of
the fractal medium, where $Z({\bf x})$ is a scalar field. 
The free energy functional has the form
\be F\{Z({\bf x})\}=F_0+\int_W {\cal F}(Z({\bf x}),\nabla Z({\bf x}))dV_3 ,\ee
where ${\cal F}(Z({\bf x}),\nabla Z({\bf x}))$ is the free energy density; 
$\nabla Z=\partial Z / \partial {\bf x}$. 
For the Ginzburg-Landau potential (\ref{2}) this density is
\[ {\cal F}(Z({\bf x}),\nabla Z({\bf x}))=\frac{1}{2} [g(\nabla Z)^2 +
aZ^2+\frac{b}{2}Z^4]. \] 
The simplest fractional generalization of (\ref{2})
can be written in the form
\be  \label{4} F\{Z({\bf x})\}=F_0+
\int_W {\cal F}(Z({\bf x}),\nabla Z({\bf x}))dV_D . \ee
Here $D$ is a fractal mass dimension of the fractal medium,
and $dV_D$ is an element of the $D$-dimensional volume:  
\be \label{5a} dV_D=C_3(D,{\bf x})dV_3. \ee
Note that the interpretation of the fractional integration
is connected with fractional dimension \cite{chaos},
which follows from the well-known formulas 
for dimensional regularizations \cite{Col}:
the fractional integral can be considered as an
integral in the fractional dimension space
up to the numerical factor $\Gamma(D/2) /( 2 \pi^{D/2} \Gamma(D))$.

For the Riesz definition of fractional integral, the 
function $C_3(D,{\bf x})$ is
\be \label{5R} C_3(D,{\bf x})=
\frac{2^{3-D}\Gamma(3/2)}{\Gamma(D/2)} |{\bf x}|^{D-3} . \ee
The initial point of the fractional integral is set to zero and
the numerical factor in Eq. (\ref{5R}) gives
the usual integral in the limit $D\rightarrow (3-0)$.
Note that the  numerical factor
$\gamma^{-1}_3(D)={\Gamma(1/2)}/{2^D \pi^{3/2} \Gamma(D/2)}$
used in Ref. \cite{SKM}
leads to $\gamma^{-1}_3(3-0)= {\Gamma(1/2)}/{2^3 \pi^{3/2}
\Gamma(3/2)}$ in the limit $D\rightarrow (3-0)$.

For the Riemann-Liouville fractional integral, 
the function $C_3(D,{\bf x})$ is 
\be \label{5RL} C_3(D,{\bf x})=
\frac{|x_1x_2x_3|^{D/3-1}}{\Gamma^3(D/3)}  . \ee
In Cartesian's coordinates $x_k$:
\[ {\bf x}=\sum^3_{k=1} x_k {\bf e}_k, \quad dV_3=dx_1 dx_2 dx_3. \]

The simplest fractional generalization of
(\ref{2}) for fractal media is
\be \label{6} F\{Z({\bf x})\}=F_0+\frac{1}{2}\int_W [g(\nabla Z)^2
+a Z^2+\frac{b}{2}Z^4] dV_D .\ee
Equation (\ref{6}) can be considered as terms of the expansion
of a functional in series over small values
$Z({\bf x})$  and the integer derivatives 
$\nabla Z=\partial Z /\partial {\bf x}$. 

For homogeneous media without external fields the parameter
$Z$ does not depend on coordinates and Eq.
(\ref{6}) gives
\be \label{7} F\{Z\}=F_0
+\frac{aV_D}{2} Z^2+\frac{bV_D}{4}Z^4 ,  \ee
where $V_D$ is a $D$-dimensional volume of fractal medium 
in the region $W$.
The equilibrium value of $Z$  corresponds to the minimum
condition for (\ref{7}):
\[ dF\{Z\}/dZ=0 . \]
If $a/b >0$, then the free energy potential has the single minimum $Z=0$.
If $a/b<0$, then there are two minima $Z=\pm \sqrt{-a/b}$.
The phase transition is realized when $a/b$ passes through zero.

\section{Ginzburg-Landau equation for fractal media}

In the general  case, the  equilibrium value of $Z({\bf x})$ 
is defined by the condition of the minimum of $F\{Z({\bf x})\}$, which
has   the   form  of variational  Euler-Lagrange  equation (\ref{1}). 
Using functional (\ref{4}) with the free energy density  
${\cal F}(Z({\bf x}),\nabla Z({\bf x}))$,  
we get the Euler-Lagrange  equation 
\be  C_3(D,{\bf x})   \frac{\partial  {\cal  F}}{\partial   Z}-
\sum^3_{k=1}\nabla_k   \Bigl(  C_3(D,{\bf x})   \frac{\partial   {\cal
F}}{\partial \nabla_k Z} \Bigr)=0 , \ee
where $C_3(D,{\bf x})$ is defined by Eq. (\ref{5R}) or (\ref{5RL}), 
and $\nabla_k = \partial / \partial x_k$, \ $k=1,2,3$.
Using (\ref{6}), we get the equation
\be \label{11} g C^{-1}_3(D,{\bf x}) \nabla_k \Bigl( C_3(D,{\bf x})
\nabla_k Z \Bigr)-aZ -bZ^3 = 0 \ee
that can be considered as the Ginzburg-Landau equation 
for fractal media in the case 
${\cal F}={\cal F}(Z({\bf x}),\nabla Z({\bf x}))$.

Eq. (\ref{11}) can be rewritten in an equivalent form
\be  g \Delta Z + E_k(D,{\bf x}) \nabla_k Z -aZ -bZ^3 = 0 ,\ee
where 
\[ E_k(D,{\bf x})=C^{-1}_3(D,{\bf x}) \nabla_k C_3(D,{\bf x}) . \]
In one-dimensional case, when $0<D \le 1$,  we have
\be \label{12-0} Z^{\prime \prime }_{xx}+\frac{D-1}{g|x|}Z^{\prime}_{x}
-(a/g)Z-(b/g)Z^3=0 , \ee
where $x=x_1$, and $Z^{\prime}_x=dZ/dx_1$. 
The simplest solutions of this equation are considered in Section 7.
Note that this equation has an analog in the form of an equation for
a nonlinear oscillator with friction. 
Therefore, the Ginzburg-Landau equation for fractal media  
describes dissipative nonlinear oscillations in coordinate space.

In the general case, the free energy  density 
can also depend on fractional derivatives of $Z$.
In this case we can  use the fractional generalization 
of the Agrawal variational equation \cite{Agrawal}.
This case will be considered in Section 6.

\section{Ginzburg-Landau equation with fractional derivatives}

Let us recall the appearance of the Ginzburg-Landau equation 
with fractional derivatives  suggested  
in Ref. \cite{Zaslavsky6}.
Consider wave propagation in some media and 
present the wave vector $\bf k$ in the form
\begin{equation}
{\bf k} = {\bf k}_0 + {\bfkappa} = {\bf k}_0 + {\bfkappa}_{\parallel}
+ {\bfkappa}_{\perp}, 
\label{eq:36}
\end{equation}
where ${\bf k}_0$ is the unperturbed wave vector and subscripts
$(\parallel ,\perp )$ are taken respectively to the direction of ${\bf k}_0$. 
Considering a symmetric dispersion law $\omega = \omega (k)$ 
for wave propagation with $\kappa \ll k_0$, we have
\begin{equation}
\omega (k) = \omega (|{\bf k}_0 + {\bfkappa}|) \approx \omega (k_0 )
+ \ c(|{\bf k}_0 + {\bfkappa}| - k_0 )
 \approx \omega (k_0 ) + c\kappa_{\parallel} +
{c\over 2k_0} \kappa_{\perp}^2 ,
 \label{eq:37}
\end{equation}
where $c = \partial\omega /\partial k_0$. 
Expression (\ref{eq:37}) in dual space ("momentum representation") 
corresponds to the following equation in coordinate space
\begin{equation}
- i {\partial Z \over \partial t} = ic {\partial Z \over \partial x_1}
  + {c\over 2k_0 } \Delta Z
   \label{eq:38}
\end{equation}
with respect to the field $Z = Z ({\bf x},t)$, where
$x_1$ is along ${\bf k}_0$.
Here we use the operator correspondence between 
the dual space and the usual space-time:
\begin{equation}
\nu \equiv \omega (k)  \ \longleftrightarrow \
 i {\partial\over\partial t} , \quad
\kappa_{\parallel} \ \longleftrightarrow \ 
 -  i {\partial\over\partial x_1} , \quad
{(\bfkappa}_{\perp})^2 \ \longleftrightarrow \ - \Delta=
- {\partial^2 \over\partial x_2 }-{\partial^2 \over\partial x_3 } , \quad
\label{eq:39}
\end{equation}
A generalization to the nonlinear case can be carried out as in (\ref{eq:37}) 
through a nonlinear dispersion law dependence on the wave
amplitude:
\begin{equation} 
\omega  =  \omega (k,|Z|^2 ) \approx\omega (k,0) + b|Z|^2
  =  \omega (|{\bf k}_0 + {\bfkappa}| ,0) + b|Z|^2
 \label{eq:40}
\end{equation}
with some constant $b = \partial\omega (k, |Z|^2 )/\partial |Z|^2$ at $|Z|^2 = 0$. 
In analogy with (\ref{eq:38}), the nonlinear equation takes the form
\begin{equation}
- i {\partial Z \over \partial t}= ic {\partial Z \over \partial x}
  + {c \over 2k_0} \Delta Z - \omega (k_0) Z - b|Z|^2 Z .
  \label{eq:41}
\end{equation}
This equation is also known as the nonlinear Schr\"{o}dinger (NLS) equation
in which any of the coefficients may be complex. Indeed, for 
$Z = Z(t,x_1-ct,x_2,x_3)$ we have
\begin{equation}
- i {\partial Z \over \partial t}
  = {c \over 2k_0} \Delta Z - \omega_0 Z - b|Z|^2 Z
  \label{eq:42}
\end{equation}
with $\omega_0 =\omega (k_0)$.

Wave propagation in media with fractal properties can be easily
generalized by rewriting the dispersion law (\ref{eq:37}). 
Namely, one can replace (\ref{eq:37}) and (\ref{eq:40}) 
by the following equation in dual space:
\begin{equation}
 \omega (k,|Z|^2 )= \omega (k_0 ,0) + c\kappa_{\parallel} +
c_{\alpha} ({\bfkappa}_{\perp}^2 )^{\alpha /2} + b|Z|^2
  \label{eq:43}
\end{equation}
with a fractional value of $1 < \alpha < 2$ and 
a new constant $c_{\alpha}$.

Using the connection between fractional derivative 
and its Fourier transform \cite{SKM} 
\begin{equation}
(-\Delta )^{\alpha /2} \longleftrightarrow ({\bfkappa}_{\perp}^2 )^{\alpha /2}
  \label{eq:44}
\end{equation}
we obtain the equation
\begin{equation}
- i {\partial Z \over \partial t} = ic {\partial Z \over \partial x}
  - {c_{\alpha} \over 2k_0} (-\Delta )^{\alpha /2}  Z - \omega_0 Z - b|Z|^2 Z,
  \label{eq:45}
\end{equation}
which corresponds to  (\ref{eq:43}) in time-coordinate space.
For $Z(t,x_1-ct,x_2,x_3)$ traveling waves as in (\ref{eq:42}), 
we get the equation
\begin{equation}
 i {\partial Z \over \partial t}
  = {c_{\alpha} \over 2k_0} (-\Delta )^{\alpha /2}  Z +  \omega_0 Z  + b|Z|^2 Z
\label{eq:46}
\end{equation}
that can be called nonstationary FGL equation or fractional nonlinear 
Schr\"{o}dinger equation (FNLS).
Let us comment on the physical structure of (\ref{eq:46}).
The first term on the right-hand side
is related to the wave propagation in a media with fractal
properties. The fractional derivative may also appear
as a result of ray chaos \cite{ref:3,ref:16}
or due to superdiffusive wave propagation (see also the discussion in
Refs. \cite{Zaslavsky2,ref:3}
and corresponding references therein). Other terms on the right-hand-side of
Eqs. (\ref{eq:45}), and (\ref{eq:46}) correspond to  wave interaction due to
the nonlinear properties of the media. Thus, Eq. (\ref{eq:46}) can describe
fractal processes of self-focusing and related issues.

We may consider a one-dimensional simplification of (\ref{eq:46}), i.e.
\begin{equation}
c {\partial Z \over \partial t}
= g {\bf D}^{\alpha}_{x} Z + aZ + b|Z|^2 Z  \label{eq:47}
\end{equation}
with some generalized constants $g, b, c$, and the Riesz fractional derivative
\begin{equation} 
({\bf D}^{\alpha}_xf)(x)=\frac{-1}{2 cos(\pi \alpha/2) \Gamma(n-\alpha)}
\frac{\partial^n}{\partial x^n} \Bigl( \int^x_{-\infty}
\frac{f(z)dz}{(x-z)^{\alpha-n+1}}+ \int^{+\infty}_x
\frac{(-1)^n  f(z)dz}{(z-x)^{\alpha-n+1}} \Bigr) , 
\label{eq:26In}
\end{equation}
where $n-1 < \alpha <n$.  
As a particular case, we can reduce Eq. (\ref{eq:47}) to  
a propagating wave solution $Z=Z(x)$, $x=x_3-ct$.
In this case, equation (\ref{eq:47}) takes the form of the stationary FGL:
\begin{equation}
g {\bf D}^{\alpha}_{x}Z+ c {\bf D}^1_{x}Z + aZ +b |Z|^2 Z = 0
\label{eq:49}
\end{equation}
or for real $Z$: 
\begin{equation}
g {\bf D}^{\alpha}_{x}Z+ c{\bf D}^1_{x}Z + aZ + bZ^3 = 0 .
\label{eq:49n}
\end{equation}
Eq. (\ref{eq:49n}) differs from   the
fractional Burgers equation \cite{ref:14,ref:13}
in the structure of the nonlinear term. Nevertheless,
an analysis similar to that in Refs. \cite{ref:14,ref:13}
may be performed to obtain some estimates on the solution.

It is well known that the nonlinear term in equations of the type
(\ref{eq:42})  leads to a steepening of the solution and its
singularity. The steepening process may be stopped by a diffusional or
dispersional term, i.e. by a higher derivative term. A similar phenomenon
may appear for the fractional nonlinear equations
(\ref{eq:47}), (\ref{eq:49}). It has been shown in Ref. \cite{ref:13}
that for the fractional Burgers equation there exists a critical value
$\alpha_c$ such that the solution is regular for all $x$
if $\alpha > \alpha_c$.

\section{Ginzburg-Landau equations with fractional derivatives in dual space}

In order to derive the FGL equations
from a variational principle, we can use two methods.
One of them  was suggested in Ref. \cite{chaos} and it will be considered here.
The second one is a generalization of 
the Agrawal variational equations  \cite{Agrawal} for the fractal media,
and will be considered in Section 6.

Let us introduce the derivatives with a fractional power of coordinates 
$x^{\beta}_k$: 
\be \label{fpd} \frac{\partial}{\partial x^{\beta}_k}=
\frac{|x_k|^{1-\beta}}{\beta} \frac{\partial}{\partial x_k}. \ee
These derivatives transform into fractional derivatives  
in dual space ("momentum representation").  
The Fourier transform of the coordinates $|x_k|^{1-\beta}$ leads us 
to the Reisz fractional derivative
\[ |x_k|^{1-\beta} \ \longleftrightarrow \ -{\bf D}^{1-\beta}_{y_k}, \]
where we use  definition  (\ref{eq:26In}).
The Fourier transform of Eq. (\ref{fpd}) is
\be \label{!!!} \frac{\partial}{\partial x^{\beta}_k}=
\frac{|x_k|^{1-\beta}}{\beta} \frac{\partial}{\partial x_k}  \
\longleftrightarrow \ -\frac{1}{\beta}{\bf D}^{1-\beta}_{y_k} y_k.  \ee


Consider free energy functional 
${\cal F}={\cal F}(Z,\partial Z/\partial x^{\beta}_k)$
for one-dimensional case. Then  the FGL equation
for  fractal media is of the form 
\be  \label{GLZ} g C^{-1}_1(\beta,x) 
\frac{\partial}{\partial x^{\beta}} \Bigl( C_1(\beta,x)
\frac{\partial}{\partial x^{\beta}} Z \Bigr)-aZ -bZ^3 = 0 ,\ee
where $C_1(\beta,x)=|x|^{\beta-1}/\Gamma(\beta)$, 
or in an equivalent form
\be \label{GLZ-2} \frac{g}{\beta^2} |x|^{2-2\beta} 
\frac{\partial^2}{\partial x^2} Z -aZ -bZ^3 = 0 . \ee
After Fourier transform of this equation, we obtain the
FGL equation in dual space:
\be  \label{GLZ-3} \frac{g}{\beta^2} {\bf D}^{2-2\beta}_y(y^2 \tilde Z) 
+a\tilde Z +b\tilde Z^3 = 0 .\ee
Here $\tilde Z(y)$ is a Fourier transform of   $Z(x)$,
i.e. $Z(x) \ \longleftrightarrow \ \tilde Z(y)$.
For the field $\Psi(y)=y^2 \tilde Z(y)$, 
we can rewrite Eq. (\ref{GLZ-3}) in the form
\[ g {\bf D}^{\alpha}_x \Psi(x)
+a(x) \Psi(x) + b(x) \Psi^3(x) = 0 , \]
where $\alpha=2-2\beta$, $0<\alpha<2$ and $a(x)=a \beta^2/|x|^2$, 
$b(x)=b \beta^2/|x|^6$.

As a result, the FGL equation for fractal media  
has fractional derivatives in dual space.
The FGL equation with fractional derivatives
in coordinate space is considered in the next section.

\section{Ginzburg-Landau equations with fractional derivatives
from fractional Agrawal principle}

Let us, first, extend the fractional Agrawal variational equations 
\cite{Agrawal} for the case of fractal media.

The simplest generalization of the free energy functional
with fractional integrals and derivatives can be written in the form
\be  \label{4b} F\{Z({\bf x})\}=F_0+
\int_W {\cal F}(Z({\bf x}),{\bf D}^{\alpha}Z({\bf x}))dV_D . \ee
Here $D$ is a fractal mass dimension of fractal medium,
and $dV_D$ is an element of $D$-dimensional volume such that
$dV_D=C_3(D,{\bf x})dV_3$. 
The function $C_3(D,{\bf x})$ is defined by 
relations (\ref{5R}) and (\ref{5RL}). 
${\bf D}^{\alpha}$ is a fractional derivative that is 
defined by Eq. (\ref{eq:26In}). The potential density is 
\be \label{FIGL} {\cal F}=\frac{1}{2} [g({\bf D}^{\alpha} Z)^2 +
aZ^2+\frac{b}{2}Z^4]. \ee

The condition of the minimum of free energy potential (\ref{4b}), (\ref{FIGL})
gives the Euler-Lagrange equation in the form
\be  C_3(D,{\bf x})   \frac{\partial  {\cal  F}}{\partial   Z}+
\sum^3_{k=1}{\bf D}^{\alpha}_{x_k}  \Bigl(  C_3(D,{\bf x})  
\frac{\partial   {\cal F}}{\partial {\bf D}^{\alpha}_{x_k} Z} \Bigr)=0 .
\label{eq36} \ee
In general, $D\not=3\alpha$ and Eq. (\ref{eq36}) gives
\be \label{F3} 
g C^{-1}_3(D,{\bf x}) \sum^3_{k=1} {\bf D}^{\alpha}_{x_k} \Bigl( C_3(D,{\bf x})
{\bf D}^{\alpha}_{x_k} Z \Bigr) + aZ + bZ^3 = 0 .
\ee
It can be considered as a fractional generalization of the 
Ginzburg-Landau equation for fractal media.
Consider some cases of this equation.

{\bf 1}. 
In the one-dimensional case $Z=Z(x)$, 
the coordinate fractional derivative is ${\bf D}^{\alpha}_x$, i.e., 
\[ {\cal F}={\cal F}(Z,{\bf D}^{\alpha}_x Z) . \]
The FGL potential density has the form
\be \label{FGL1} {\cal F}=\frac{1}{2} g
\Bigl({\bf D}^{\alpha}_x Z \Bigr)^2 +
\frac{a}{2}Z^2+\frac{b}{4}Z^4. \ee 
Using the formulas for fractional integration by parts 
\[ \int_{-\infty}^{\infty} f(x) \frac{d^{\beta} g(x)}{d x^{\beta}} dx=
\int_{-\infty}^{\infty} g(x) \frac{d^{\beta} f(x)}{d (-x)^{\beta}} dx , \]
\be \label{ibp} \int_{-\infty}^{\infty} f(x) {\bf D}^{\alpha}_x g(x)dx =
\int_{-\infty}^{\infty} g(x){\bf D}^{\alpha}_{x} f(x) dx , \ee
we obtain the Euler-Lagrange equation
\be  {\bf D}^{\alpha}_x   \Bigl(  C_1(D,x)   
\frac{\partial   {\cal F}}{\partial {\bf D}^{\alpha}_x Z} \Bigr)+ 
C_1(D,x) \frac{\partial {\cal F}}{\partial Z}=0 , \ee
where the function $C_1(D,x)$ is defined as
\be \label{C1} C_1(D,x)=\frac{|x|^{D-1}}{\Gamma(D)} . \ee
Using for ${\cal F}$ (\ref{FGL1}), we arrive at
\be  C^{-1}_1(D,x) {\bf D}^{\alpha}_x   \Bigl(  C_1(D,x)   
{\bf D}^{\alpha}_x Z  \Bigr)+aZ+bZ^3 =0 . \ee
For the case $D=1$, we have $C_1=1$ and (\ref{C1}) transforms into
\begin{equation}
{\bf D}^{2\alpha}_{x}Z+ aZ + bZ^3 = 0,
\label{eq:43n}
\end{equation}
where ${\bf D}^{\alpha}_x$ is the Riesz derivative.

{\bf 2}.
The free energy functional density ${\cal F}$ can be  
a function of  $Z=Z(x)$, and
two different fractional derivatives ${\bf D}^{\alpha}_x Z$, 
and ${\bf D}^{\beta}_x Z$, $\alpha\not=\beta$, i.e., 
\[ {\cal F}={\cal F}(Z,{\bf D}^{\alpha}_x Z,{\bf D}^{\beta}_x Z) . \]
The FGL potential density has the form
\be \label{FGL2} {\cal F}=\frac{1}{2} g_1
({\bf D}^{\alpha}_x Z)^2 +
\frac{1}{2}g_2 ({\bf D}^{\beta}_x Z)^2+\frac{a}{2}Z^2+
\frac{b}{4}Z^4. \ee 
For example, $\alpha=1$, $0<\beta<1$. Using again (\ref{ibp}), 
we obtain the following Euler-Lagrange equation,  
\be  {\bf D}^{\alpha}_x   \Bigl(  C_1(D,x)   
\frac{\partial   {\cal F}}{\partial {\bf D}^{\alpha}_x Z} \Bigr)
+{\bf D}^{\beta}_x \Bigl(  C_1(D,x)   
\frac{\partial   {\cal F}}{\partial {\bf D}^{\beta}_x Z} \Bigr)+
C_1(D,x)   \frac{\partial  {\cal  F}}{\partial   Z}=0 , \ee
where $C_1(D,x)$ is given in (\ref{C1}). 
For the FGL  density (\ref{FGL2}), we get
the FGL equation for fractal media
\be  g_1C^{-1}_1(D,x)  {\bf D}^{\alpha}_x  \Bigl(  C_1(D,x)   
{\bf D}^{\alpha}_x Z  \Bigr)+
g_2 C^{-1}_1(D,x)  {\bf D}^{\beta}_x  \Bigl(  C_1(D,x)   
{\bf D}^{\beta}_x Z \Bigr)+aZ+bZ^3 =0 . \ee
Particularly, when $D=1$ and $C_1=1$, we have
\[ g_1 {\bf D}^{2\alpha}_{x}Z+ g_2 {\bf D}^{2\beta}_{x}Z+ aZ + bZ^3 = 0, \ \
0 \le \alpha , \beta \le 1. \]

{\bf 3}.
In  the general case, the free energy density  functional is a function 
of $Z=Z({\bf x})$, and derivatives ${\bf D}^{\alpha_k}_{x_k} Z$ of 
fractional orders $\alpha_k$ different for coordinates $x_k$,  i.e., 
\be \label{F2} 
{\cal F}={\cal F}(Z,{\bf D}^{\alpha_1}_{x_1} Z,
{\bf D}^{\alpha_2}_{x_2} Z,{\bf D}^{\alpha_3}_{x_3} Z) . \ee
For this case, the FGL equation for fractal media is
\be g C^{-1}_3(D,{\bf x}) \sum^3_{k=1} {\bf D}^{\alpha_k}_{x_k} 
\Bigl(  C_3(D,{\bf x})  {\bf D}^{\alpha_k}_{x_k} Z  \Bigr)+aZ+bZ^3 =0 . \ee
The sum of orders $\alpha_k$ can be equal to the fractal mass dimension $D$
of the medium
\[ \alpha_1+\alpha_2+\alpha_3=D, \]
but in the general case it can be that
$\alpha_1+\alpha_2+\alpha_3\not=D$.

\section{Simplest solutions}

Let us consider a simple example of the Ginzburg-Landau equation (\ref{11})
for fractal media in the one-dimentional case. Then
\[ \frac{\partial Z}{\partial x_2}=0, \quad \frac{\partial Z}{\partial x_3}=0, \]
and $Z=Z(x_1)$. Eq. (\ref{GLZ}) for one-dimensional case
$\beta_1=D$ and $0<D<1$ transforms into
\be  \label{11b} g C^{-1}_1(D,x) \frac{\partial}{\partial x}
\Bigl( C_1(D,x) \frac{\partial}{\partial x} Z \Bigr)-
aZ -bZ^3 = 0 ,\ee
where $x=x_1$, and $C_1(D,x)$ is defined by (\ref{C1}). 
This equation can be rewritten in an equivalent form
\be \label{12} Z^{\prime \prime }_{xx}+\frac{D-1}{g|x|}Z^{\prime
}_{x}-(a/g)Z-(b/g)Z^3=0 , \ee
where $Z^{\prime}_x=dZ(x)/dx$. 
Note that Eq. (\ref{12}) has an analog of the equation for a
nonlinear oscillator with friction
\[ \ddot{x}(t)=\alpha x(t)+\beta x^3(t)-\gamma(t) \dot{x}(t), \]
where we use
\[ \alpha=a/g, \quad \beta=b/g, \quad \gamma(t)=(D-1)/g|t| . \]
Therefore the FGL equation for fractal media (\ref{11b}) 
describes coordinate nonlinear oscillations with a ``dissipative''-like term.

Let us consider  solution of  (\ref{11b}) with $b=0$. 
$Z({\bf x})$ satisfies the equation
\[ g[C_1(D,x) Z^{\prime}_x]^{\prime}_x-a C_1(D,x) Z=0 , \]
where $x\in(0;\infty)$, that can be rewritten as
\be \label{11c} g x Z^{\prime \prime}_{xx}(x)+(D-1)Z^{\prime}_{x}(x)-
a x Z(x)=0 . \ee
The corresponding solution is
\[ Z(x)=C_1x^{1-D/2}J_{\nu}(\sqrt{-a/g} x)+
C_2 x^{1-D/2} Y_{\nu}(\sqrt{-a/g} x) , \]
where $\nu= |1-D/2|$, $J_{\nu}(x)$, and $Y_{\nu}(x)$ 
are the Bessel functions of the first and second kind.

The solution of Eq. (\ref{11c}) with the conditions
\[ Z(1)=e, \quad Z^{\prime}(1)=c \]
has the form \\ 
\[ Z(x)=x^{1-D/2} J_{D/2-1}(\sqrt{-A}x)  
\Bigl(cD\sqrt{-A}\ Y_{D/2}(\sqrt{-A})-(eA)\sqrt{-A}\ 
Y_{D/2}(\sqrt{-A})+ \]
\[ +cA\ Y_{D/2+1}(\sqrt{-A})\Bigr) / 
\Bigl( A\sqrt{-A} (\ J_{D/2+1}(\sqrt{-A})\ Y_{D/2}(\sqrt{-A})-
\ Y_{D/2+1}(\sqrt{-A})\ J_{D/2}(\sqrt{-A}))  \Bigr)+\]

\[ +x^{1-D/2} Y_{D/2-1}(\sqrt{-A}x) 
\Bigl(-cD\ J_{D/2}(\sqrt{-A})+\sqrt{-A}c \ J_{D/2+1}(\sqrt{-A})+ \]
\[ +eA\ J_{D/2}(\sqrt{-A})\Bigr)/
\Bigl( A (\ J_{D/2+1}(\sqrt{-A})\ Y_{D/2}(\sqrt{-A})-
\ Y_{D/2+1}(\sqrt{-A})\ J_{D/2}(\sqrt{-A})) \Bigr) \] 
where $A=a/g$.


\begin{figure}
\centering
\rotatebox{270}{\includegraphics[width=9 cm,height=9 cm]{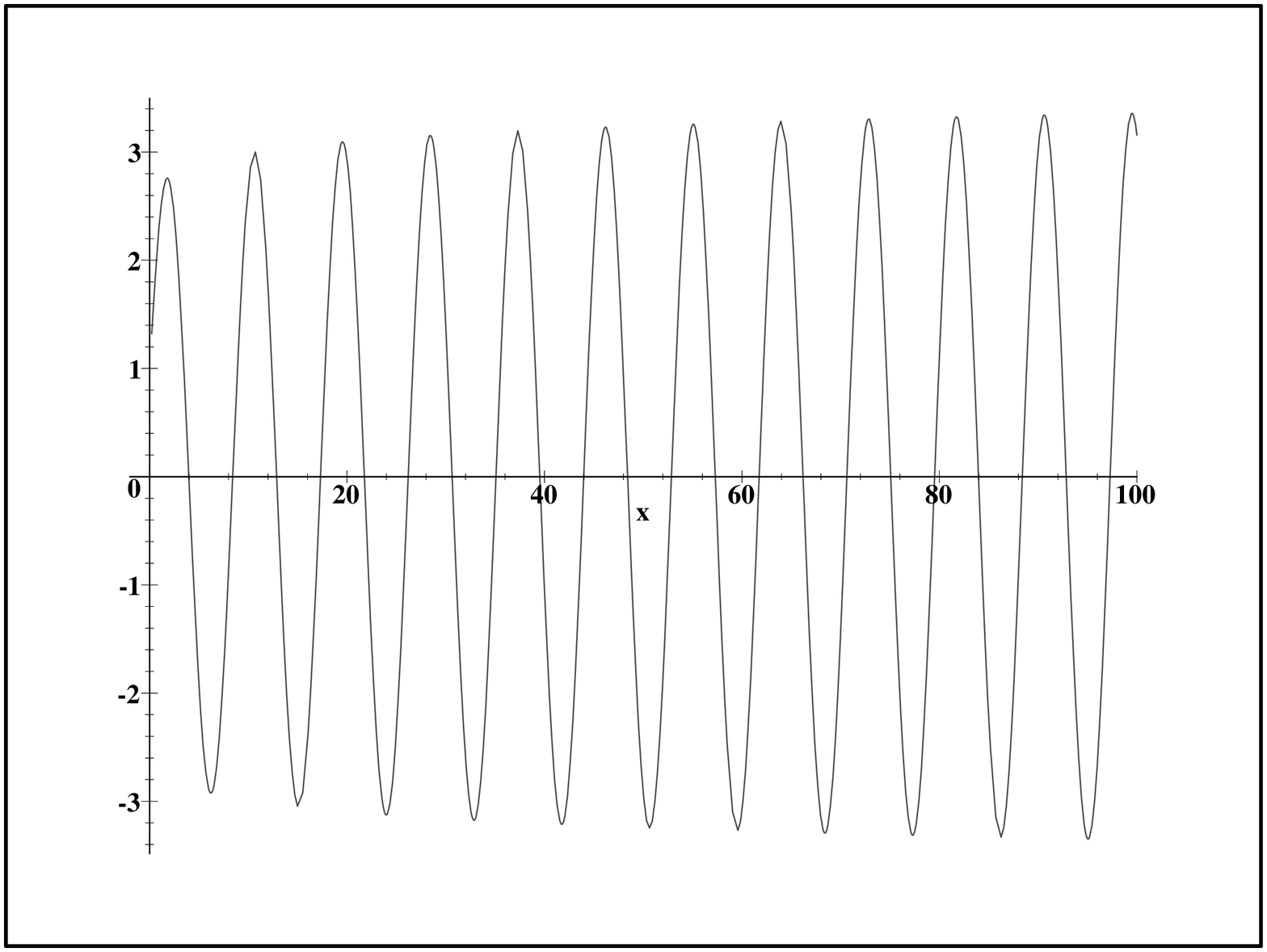}}
\caption{\it Solution for $b=0$, $a/g=-1$ and $D=9/10$.}
\label{gl9-10}
\end{figure}

\begin{figure}
\centering
\rotatebox{270}{\includegraphics[width=9 cm,height=9 cm]{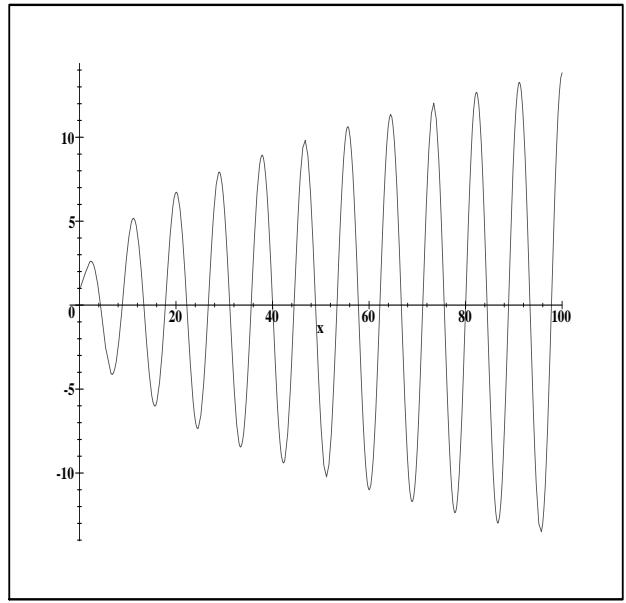}}
\caption{\it Solution for $b=0$, $a/g=-1$ and $D=1/10$.}
\label{gl1-10}
\end{figure}

The solutions  with $b=0$, $g=1$, and $a/g=-1$
are demonstrated in Fig. 1 for $D=9/10$ and in Fig. 2 for $D=1/10$.
The growth of solution is saturated by the nonlinear term and full
solution for  (\ref{12}) may appear similar to the vicinity of a stable
limit cycle.

\section{Conclusion}

Derivatives and integrals of fractional order have found many
applications in recent studies in physics.
In this paper, we use the fractional integrals to describe fractal media 
with noninteger mass dimensions.
Fractional integration can also be used to describe time dependent 
processes in fractal media.
The fractional Ginzburg-Landau equation for fractal media is derived from
the corresponding generalization of free energy functional and
variational Euler-Lagrange equations. 
Generalization of the Agrawal variational equation 
\cite{Agrawal} appears to be very useful for this goal.

The fractional continuous models have wide applications.
Partly, it is because  relatively small number of parameters
can define a complex medium with a rich structure.
In many cases, the real fractal structure of matter
can be disregarded and the medium can be described by
some fractional continuous mathematical model.
Smoothing of the microscopic characteristics over the
physically infinitesimal volume transform the initial
fractal medium into a fractional continuous model
that uses the fractional integrals of the same order as
the fractal mass dimension of the media.
The fractional integrals can be considered as an approximation
of the integrals on fractals \cite{RLWQ}. Note that
fractional integrals can be considered as integrals
over the space with fractional dimension up 
to a numerical factor \cite{chaos}.
Some applications of the FGL equation 
was discussed also in Ref. \cite{Milovanov}.

\section*{Acknowledgment}

G.M.Z. was 
supported by the Office of Naval Research, Grants No. N00014-02-1-0056,
U.S. Department of Energy Grant No. DE-FG02-92ER54184, and the NSF Grant 
No. DMS-0417800.



\end{document}